\documentclass[fleqn,twoside]{article}
\usepackage{amsmath}
\usepackage{espcrc2}

\usepackage{graphicx}
\usepackage{psfrag}

\newcommand{\lt}{\left}
\newcommand{\rt}{\right}

\newcommand{\Dsp}{\not\!\! D_\perp}
\newcommand{\Dp}{D_\perp}
\newcommand{\vv}{\vec{v}}
\newcommand{\vD}{\vec{D}}
\newcommand{\cL}{{\cal L}}
\newcommand{\order}{{\cal O}}
\newcommand{\bean}{\begin{eqnarray}}
\newcommand{\eean}{\end{eqnarray}}
\newcommand{\KK}{\vv\cdot\vD}
\newcommand{\vBp}{\vec{B'}}
\newcommand{\vE}{\vec{E}}
\newcommand{\vB}{\vec{B}}
\newcommand{\vEp}{\vec{E'}}
\newcommand{\be}{\begin{equation}}
\newcommand{\ee}{\end{equation}}

\title{Moving NRQCD for B Form Factors at High Recoil}

\author{Kerryann M. Foley\address[Cornell]{Laboratory for Elementary-Particle Physics, Cornell University, Ithaca, NY 14853}, G. Peter Lepage\addressmark[Cornell]}
   
\begin{document}

\begin{abstract}
We derive the continuum and lattice tree-level moving NRQCD (mNRQCD) through order  $1/m^2$.
mNRQCD is a generalization of NRQCD for dealing with hadrons with nonzero velocity $u_\mu$.
The quark's total momentum is written as $P^\mu=Mu^\mu+k^\mu$ where $k^\mu\ll Mu^\mu$ is discretized and $Mu^\mu$ is treated exactly.
Radiative corrections to couplings on the lattice are discussed. 
mNRQCD is particularly useful for calculating $B\rightarrow\pi$ and $B\rightarrow D$ form factors since errors are similar at low and high recoil. 
\vspace{1pc}
\end{abstract}

\maketitle

\section{INTRODUCTION}
Decays of heavy quarks are essential for determining the CKM elements.  
Because of the mass difference between the initial and final states, these decays can lead to large recoil momenta.
Current lattice simulations require that the inverse lattice spacing, $a^{-1}$, be greater then the largest momenta to keep errors reasonable.
For the important case $B\rightarrow \pi l\nu_l$, needed for $V_{ub}$, the high recoil limit gives $p_\pi\rightarrow 2.5$ GeV. This implies $a^{-1}\gg 2.5$ GeV, which is much  more then current simulations can handle.

A solution to this problem has two parts.
First, we use a moving B meson and so share the momentum between the $B$ and the $\pi$.
This allows us to find a frame in which all light quark momenta are the same order and errors are minimized.
Second, defining the momentum of the heavy quark as
\be P^\mu_b=M_bu^\mu+k^\mu,\label{eq:p} \ee
where $u^\mu$ is the 4-velocity of the B meson, means that $k^\mu\sim\lt(\frac{\Lambda}{M}\rt)Mu^\mu\ll Mu^\mu$.
We then discretize $k^\mu$ while treating $M_bu^\mu$ exactly.
This removes the large, uninteresting part of the $b$'s momentum and allows larger lattice spacings to be used. 
This procedure is analogous to NRQCD where the heavy, stationary quark has momentum $P^\mu=M+k^\mu$ and $k^\mu$ is discretized.

\section{DECAY KINEMATICS}
The $B$ meson momentum, $P_B=\lt(P,M_B^2/P\rt)$, can be split between the b quark and the spectator quark in its wavefunction:

\psfrag{C}[lt]{B}
\psfrag{V}[rt]{}
\psfrag{D}{u}
\psfrag{W}{$\;\;P_b=(1-x)P_B-p_\perp$,}
\psfrag{E}{b}
\psfrag{X}{$\;\;P_u=xP_B+p_\perp$}
\includegraphics[scale=0.6]{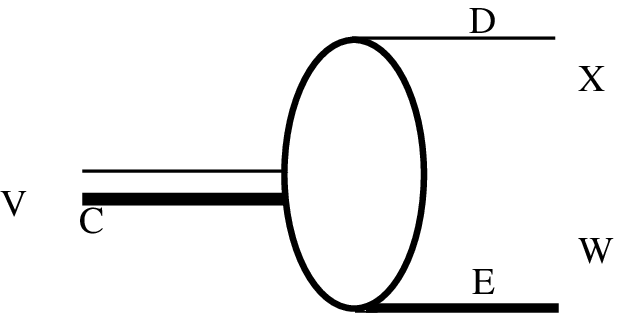}
\vspace{1pc}

\noindent where the choice of $P$ describes the choice of frame.

The light quark has transverse momentum $p_\perp\sim\Lambda\sim 500$ MeV with $\Delta p_\perp\sim\Lambda$, and its momentum fraction is peaked at $x\sim\frac{\Lambda}{M_b}$ with range $\Delta x\sim\frac{\Lambda}{M_b}$.
Consequently, 
\be k\sim P_u\sim \Lambda u \ee
which implies that discretization errors associated with the heavy quark match those of the light quark.
Note also that $u_B\cdot k\sim\Lambda$ even as $u_B^\mu$, $k^\mu\rightarrow\infty$.

Our goal is to find a frame, for the simulation,  where the discretization errors are the same for all quarks. 
To see how the choice of frame affects our calculation we give three examples for $B\rightarrow \pi l\nu$ in the high-recoil limit.
In the $B$'s rest frame, $P=M_B$, and
\be P_\pi\sim\frac{M_B}{2},\;\;\;\;\;k\sim P_u\sim\Lambda\;\;\;\Rightarrow\;\;\; P_\pi\gg P_u. \ee
With this choice discretization errors associated with the pion are far larger than other discretization errors.
In the $\pi$ rest frame, $P=m_\pi$, and
\be P_\pi\sim\Lambda\;\;\;\;\;k\sim P_u\sim\frac{\Lambda M_B}{2M_\pi}\;\;\;\Rightarrow\;\;\;P_\pi\ll P_u; \ee
the $B$'s discretization errors dominate.
The best frame is the ``spectator-$\pi$ Breit'' frame,  $P=\sqrt{\Lambda M_B}$, where
\be P_\pi\sim\frac{\sqrt{\Lambda M_B}}{2}\;\;\;\;\;k\sim P_u\sim\frac{\sqrt{\Lambda M_B}}{2}\label{eq:Breit} \ee
\be\;\;\;\Rightarrow\;\;\;P_\pi\sim P_u\sim k\sim 800{\:\rm MeV}\nonumber \ee
In this frame the errors are distributed evenly between both hadrons, and an inverse lattice spacing $a^{-1}\sim 2$ GeV is adequate.
This is a savings of $\sqrt{\frac{\Lambda}{M_B}}\sim\frac{1}{3}$ compared to the B rest frame making simulations $1000$ times faster.

\section{mNRQCD DERIVATION}
To derive the tree-level effective Lagrangian for a moving $b$ quark we use the FWT method to transform the QCD Lagrangian order by order in $1/m$ into non-interacting quark and anti-quark pieces.
Working through order $1/m^2$, this gives the well known HQET result
\bean\cL\!\!\!&=&\!\!\overline{Q}\lt(iu\cdot D-\frac{\Dp^2}{2m}-\frac{g}{4m}F_{\mu\nu}\sigma^{\mu\nu}\rt.\nonumber\\
&&\!\!\!\!\!-\frac{g}{8m^2}u^\alpha\lt[\Dp^\nu F_{\alpha\nu}\rt]\nonumber\\
&&\!\!\!\!\!\lt.+\frac{ig}{8m^2}u^\alpha\sigma^{\mu\nu}\lt\{D_\mu,F_{\alpha\nu}\rt\}+\order\lt(\frac{\Lambda^3}{m^3}\rt)\rt)Q\label{eq:HQET} \eean
and the original wave function $\psi$ can be written in terms of $Q$.

To implement this formalism efficiently on a Euclidean lattice, we must convert it to an initial value problem (in time). 
To do this we first reduce the above equations to a 2-component form and then remove higher order time derivatives using a field redefinition.

To recast the Lagrangian into 2-component form we write $Q$ as a rotation of a spinor with two zero components:
\bean Q\!\!\!&=&\!\!\!\Lambda\lt(\!\begin{array}{c}\chi\\0\end{array}\!\rt)\nonumber\\
\!\!\!&=&\!\!\!\!\frac{1}{\sqrt{2\lt(1+\gamma\rt)}}\!\lt(\!\begin{array}{cc}1+\gamma&\!\gamma\lt(\vec{\sigma}\cdot\vv\rt)\\ \gamma\lt(\vec{\sigma}\cdot\vv\rt)&\!1+\gamma\end{array}\!\rt)\!\!\lt(\!\begin{array}{c}\chi\\0\end{array}\!\rt)\eean
Higher order time derivatives are removed using a field redefinition $A_{{\rm FR}}$,
\be \chi\rightarrow A_{{\rm FR}}\chi\;\;\;\chi^\dag\rightarrow \chi^\dag A_{{\rm FR}} \ee
The resulting mNRQCD to order $\lt(\Lambda/m\rt)^2$ is:
\bean\cL\!\!\!&=&\!\!\!\chi^\dag\lt( iD_t+i\KK\rt.\label{eq:Lcontinuum}\\
\!\!\!&&\!\!\!+\frac{1}{2m\gamma}\lt(\vD^2-\lt(\KK\rt)^2+\sigma\cdot\vBp\rt)\nonumber\\
\!\!\!&&\!\!\!+\frac{1}{8m^2}\lt[\vD\cdot \vE\rt]-\frac{\vv^2}{16m^2}\vv\cdot\lt[\vD\times \vB\rt]\nonumber\\
\!\!\!&&\!\!\!+\frac{i}{4m^2\gamma^2}\lt(\lt\{\KK,\vD^2\rt\}-2\lt(\KK\rt)^3\rt)\nonumber\\
\!\!\!&&\!\!\!-\frac{\lt(2-\vv^2\rt)}{16m^2}\lt[\KK,v\cdot \vE\rt]\nonumber\\
\!\!\!&&\!\!\!+\frac{i}{8m^2\gamma}\sigma\cdot\lt(\vD\times \vEp-\vEp\times\vD\rt)\nonumber\\
\!\!\!&&\!\!\!+\frac{i}{8m^2\lt(\gamma+1\rt)}\sigma\cdot\lt\{\KK,\vv\times \vEp\rt\}\nonumber\\
\!\!\!&&\!\!\!\lt.+\frac{i}{4m^2}\lt\{\KK,\sigma\cdot\vBp\rt\}+\order\lt(\frac{1}{m^3}\rt)\rt) \chi\nonumber \eean
where $u^\mu=\lt(\gamma,\gamma\vv\rt)$ and $\vEp$ and $\vBp$ are the electric and magnetic fields in the $b$ quark's rest frame:
\be\vec{E}'=\gamma\lt(\vE+\vv\times\vec{B}-\frac{\gamma}{\gamma+1}\vv\lt(\vv\cdot\vE\rt)\rt), \ee
\be\vec{B}'=\gamma\lt(\vec{B}-\vv\times\vec{E}-\frac{\gamma}{\gamma+1}\vv\lt(\vv\cdot\vec{B}\rt)\rt). \ee
mNRQCD reduces to NRQCD when $\vv\rightarrow 0$.
Lower order versions of this formalism are considered in~\cite{Sloan:1997fc,Hashimoto:1995in,Mandula:1997hb}.

Currents and other operators involving the $b$ quark can be designed using the b quark field:
\bean \lefteqn{\psi \lt(x\rt)\equiv\frac{e^{-imu\cdot x}}{\sqrt{2\gamma\lt(1+\gamma\rt)}}}\\
\lefteqn{\;\;\;\;\;\;\;\;\;\times\lt(1+\frac{i\Dsp}{2m}+\frac{1}{8m^2}\lt(2u\cdot D\Dsp-\Dsp^2\rt)\rt)}\nonumber\\
\lefteqn{\;\;\;\;\;\;\;\;\;\times\rm{A_{FR}}\:\Lambda\lt(\begin{array}{c}\chi\\0\end{array}\rt)+\order\lt(\frac{\chi}{m^3}\rt)}\nonumber \eean

Note that our asymmetric treatment of time and space means that our Lagrangian relies upon an expansion in $\vec{u}=\gamma\vv$ rather then $\Lambda/m$ as in HQET.
Since $|\vec{u}|\rightarrow\infty$ as $\vv\rightarrow 1$ this could be problematic.
Our choice of frame for $B\rightarrow\pi l\nu$, however, means that $\gamma\vv$ in never larger then $0.35$ and the truncation errors are no larger then the $1/m^3$ terms neglected in~\eqref{eq:HQET}.

\section{Lattice mNRQCD}
We reformulate the mNRQCD Lagrangian of~\eqref{eq:Lcontinuum} on the lattice using the usual forms for lattice derivatives and fields as in~\cite{Lepage:1992tx}.
To achieve high precision, we must include perturbative corrections to our tree-level coefficients.
Renormalizations through order $\alpha_s^2$ are needed for the leading order $\vv\cdot\vD$:
\be\chi^\dag\lt(D_t -i\KK\rt)\chi\nonumber \ee
\be\rightarrow\chi^\dag\lt(D_t-i\xi\KK-i\xi_3 \frac{a^2}{6}v_iD_i^3\rt)\chi \ee
where $\xi=1+\xi_1\alpha_s+\xi_2\alpha_s^2$

For $D$ mesons, where $\frac{\Lambda}{M}\sim \frac{1}{3}$, we need to include terms of order $\frac{1}{m^2}$ through tree-level and $\frac{1}{m}$ through order $\alpha_s$.
This requires 5 additional renomalizations
\bean\lefteqn{\chi^\dag\lt(\vD^2-\lt(\KK\rt)^2+\gamma\sigma\cdot\vBp+c_3\gamma\sigma\cdot\vB\rt)\chi}\nonumber\\
\lefteqn{\rightarrow\;\chi^\dag\lt(c_1\vD^2-c_2\lt(\KK\rt)^2\rt)\chi}\\
\lefteqn{\;\;+\chi^\dag\lt(-c_4\gamma\sigma\cdot\vv\times\vE+c_5\frac{\gamma^2}{\gamma+1}\sigma\cdot\vv\lt(\vv\cdot\vB\rt)\rt)\chi}\nonumber \eean
where $c_i=1+c_{i,1}\alpha_s$.
For B mesons, where $\frac{\Lambda}{M}\sim\frac{1}{10}$, we need $\frac{1}{m}$ only through tree-level.
The calculation of the necessary perturbative coefficients is currently in progress.

\section{Conclusion}
The method shown here is easily extensible to other processes with large recoil including $B\rightarrow Dl\nu$ which is used to find $V_{cb}$, $D\rightarrow \pi l\nu$ used for $V_{cd}$ and $B\rightarrow K^*\gamma$ 
Using mNRQCD, errors are comparable at at both low and high recoil.
We can vary the momentum transfer  continuously by varying the B meson velocity, $u_B$. 
\subsection*{Acknowledgment}
This material is based upon work supported by the National Science Foundation under Grant No. PHY-0098631.

\end{document}